\documentclass{article}

\usepackage{microtype}
\usepackage{graphicx}
\usepackage{subfigure}
\usepackage{booktabs} 
\usepackage[T1]{fontenc}  
\usepackage[utf8]{inputenc}

\usepackage{hyperref}

\usepackage[accepted]{icml2020}

\icmltitlerunning{I know why you like this movie: Interpretable Efficient Mulitmodal Recommender}

\begin{document}

\twocolumn[
\icmltitle{I know why you like this movie: Interpretable Efficient Mulitmodal Recommender}



\icmlsetsymbol{equal}{*}

\begin{icmlauthorlist}
\icmlauthor{Barbara Rychalska}{synerise,pw}
\icmlauthor{Dominika Basaj}{synerise,pw}
\icmlauthor{Jacek~Dąbrowski}{synerise}
\icmlauthor{Michał Daniluk}{synerise}
\end{icmlauthorlist}


\icmlaffiliation{synerise}{Synerise}
\icmlaffiliation{pw}{Warsaw University of Technology}

\icmlcorrespondingauthor{Barbara Rychalska}{barbara.rychalska@synerise.com}

\icmlkeywords{Machine Learning, ICML}

\vskip 0.3in
]



\printAffiliationsAndNotice{} 

\begin{abstract}
Recently, the Efficient Manifold Density Estimator (EMDE) model has been introduced. The model exploits Local Sensitive Hashing and Count-Min Sketch algorithms, combining them with a neural network to achieve state-of-the-art results on multiple recommender datasets. However, this model ingests a compressed joint representation of all input items for each user/session, so calculating attributions for separate items via gradient-based methods seems not applicable. We prove that interpreting this model in a white-box setting is possible thanks to the properties of EMDE item retrieval method. By exploiting multimodal flexibility of this model, we obtain meaningful results showing the influence of multiple modalities: text, categorical features, and images, on movie recommendation output.

\end{abstract}

\section{Introduction}

Most recommender systems are hard to intepret as they focus just on user interactions, and the connection between the user sentiment and various input modalities such as movie plot or visual style cannot be understood directly \cite{10.1145/3308558.3313710}. In  content-based or hybrid recommenders which ingest multimodal data, the focus on intepretability may enforce the application of a specific interpretable architecture at the cost of performance \cite{ijcai2018-472}. Interpreting the decisions of a recommender system is a very important business task, allowing to discover crucial factors explaining which input features increase or decrease the attractiveness of an item. The problem grows even more important as traditional performance measures such as Recall@N do not directly translate into user satisfaction \cite{10.1145/2209310.2209314}. Multimodal intepretability can be a step in the direction of creating better performance measures. 

In this paper, we aim to interpret the decisions made by the state-of-the-art EMDE model \cite{emde}, which allows us to leverage a wide array of diverse data: liked and disliked movies (interaction data), movie plot (text), movie cast (semi-categorical data), and movie posters (image). To the best of our knowledge, no other recommender system gives the chance to interpret the influence of so many diverse modalities. We find a solution to the problem of deriving attributions for individual items even though the input to EMDE is the \textit{sketch}: a joint compressed representation of items. Interestingly, we find that the system learns to extract relevant features of image and text inputs based on just pure interaction output. Some discovered influential factors are not easily understandable and demand a deeper analysis. As such, our results add value beyond what might be guessed by human intuition alone.


\begin{figure*}
\centering
\subfigure[]{
    \label{fig:a}
    \includegraphics[scale=0.25]{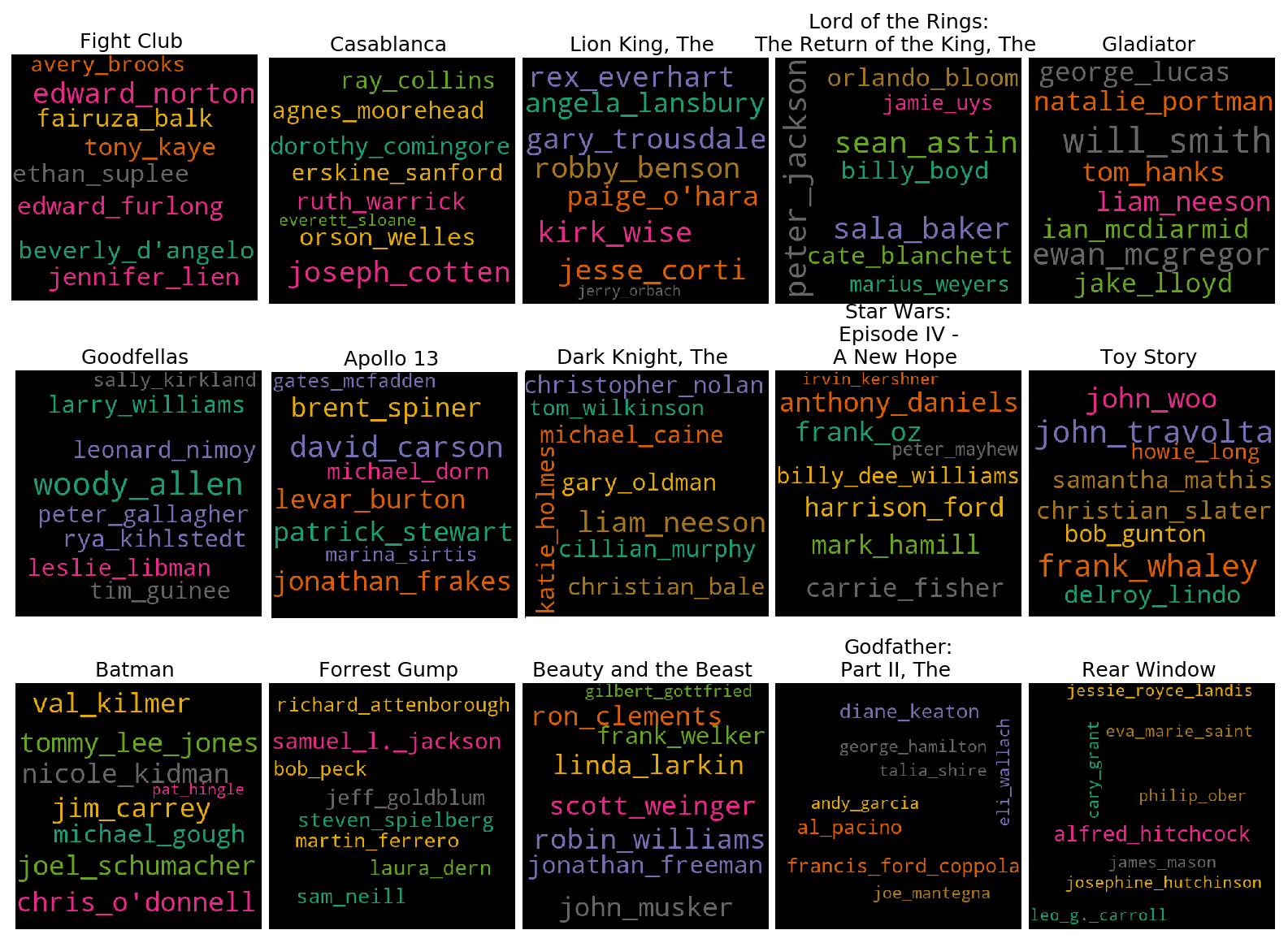}
    }
\hspace{-1\baselineskip}
\subfigure[]{\label{fig:a}\includegraphics[scale=0.25]{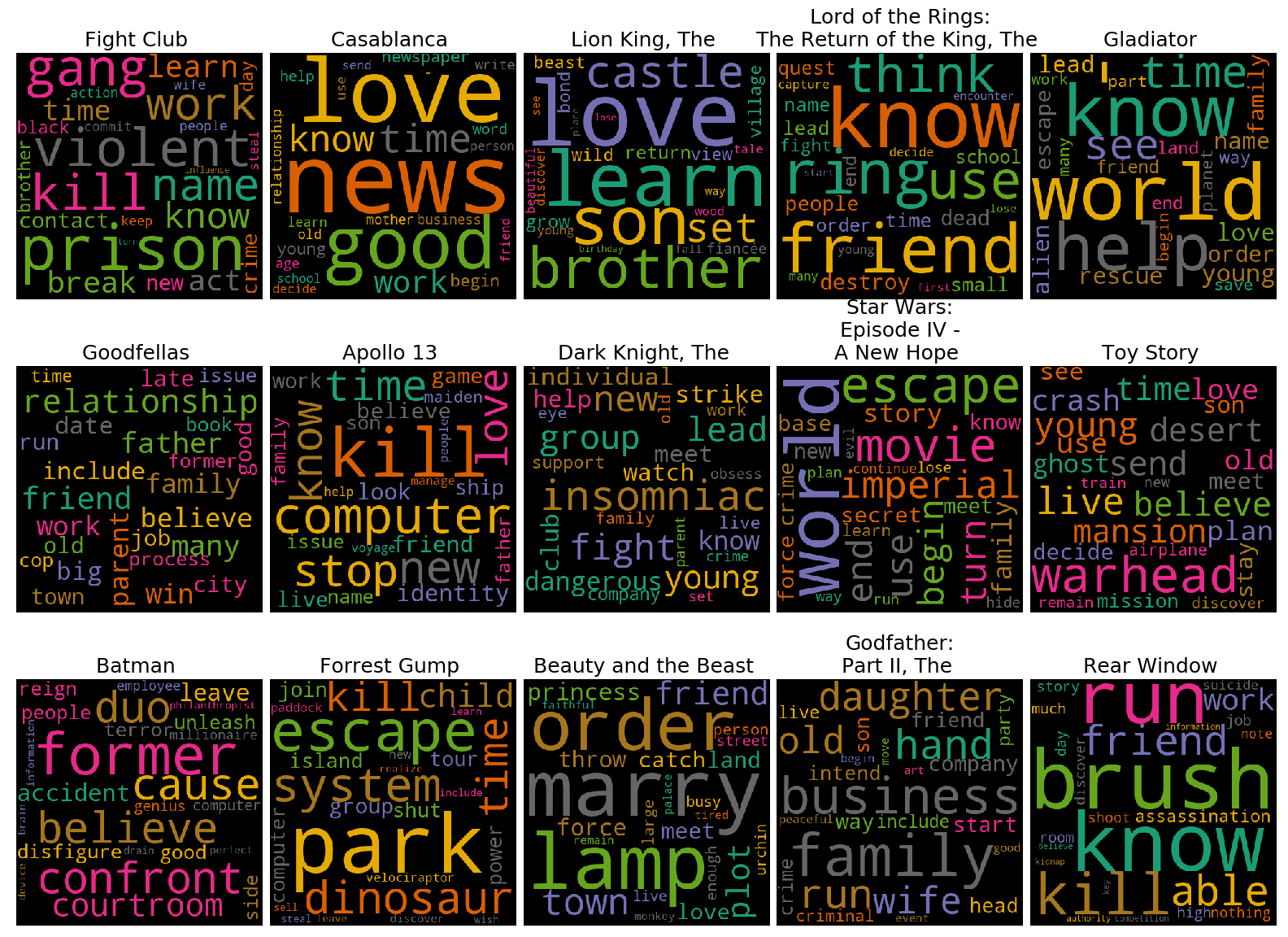}}
\hspace{-1\baselineskip}
\caption{Image a) shows tokens from move descriptions, and b) shows cast with the strongest positive attribution to output movies. Token size reflects the attribution score.}
\label{cast-desc}
\end{figure*}

\begin{figure*}%
\centering
\includegraphics[scale=0.33]{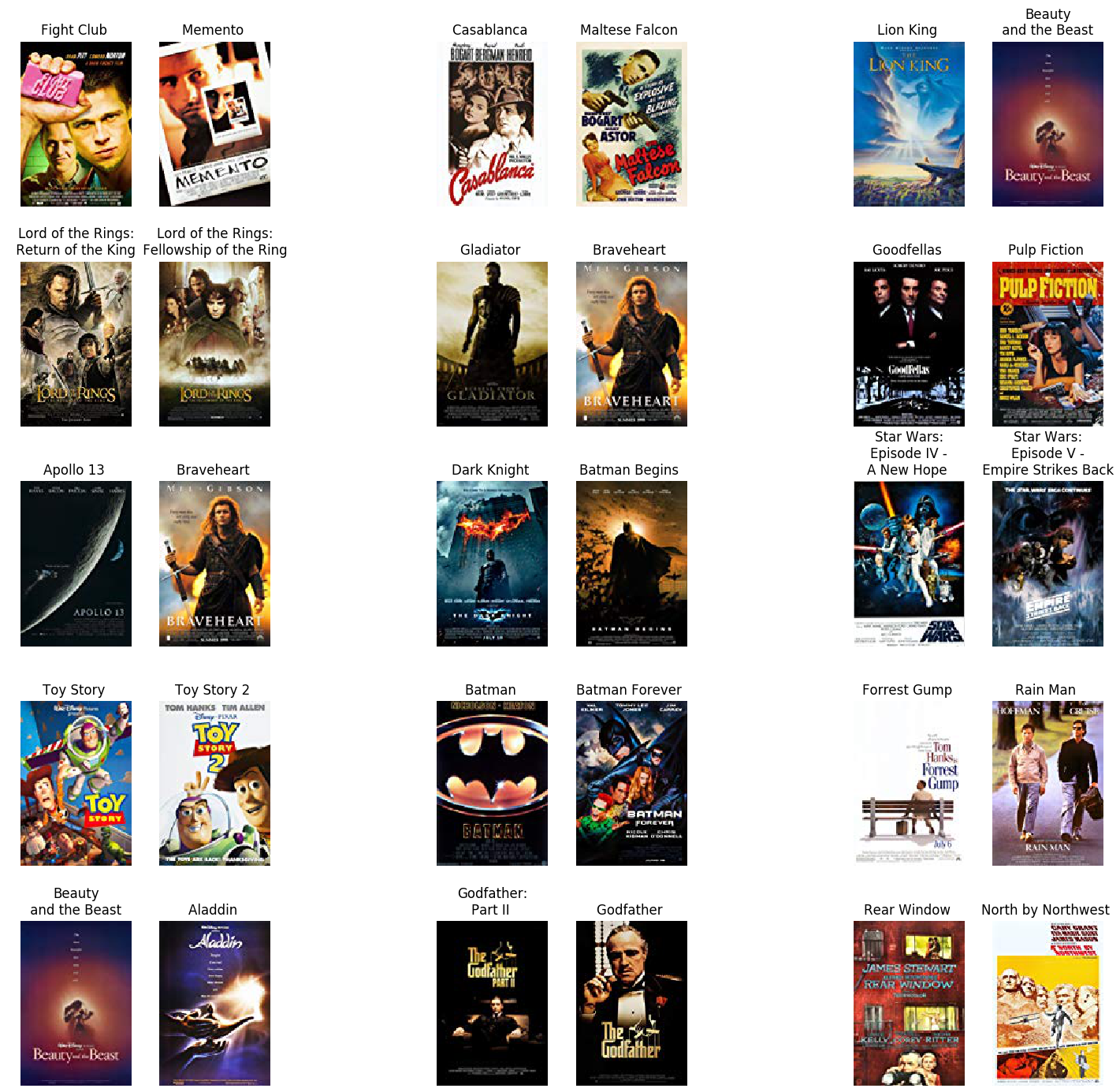}
\caption{Interpretability of posters. The left-hand side poster is the recommended movie poster, while the right-hand side poster is the most influential poster for this recommendation. Some cases are easily understandable, e.g. the pair \textit{Gladiator} and \textit{Braveheart} - the key influencer is the exposition of the main character in a statue-like pose towering over the observer (with the head starting just at the top of the poster). Some examples such as the pair \textit{Apollo 13} and \textit{Braveheart}, are less clear to interpret.}
\label{posters}
\end{figure*}

\begin{figure*}%
\centering
\subfigure[]{\label{fig:a}\includegraphics[scale=0.25]{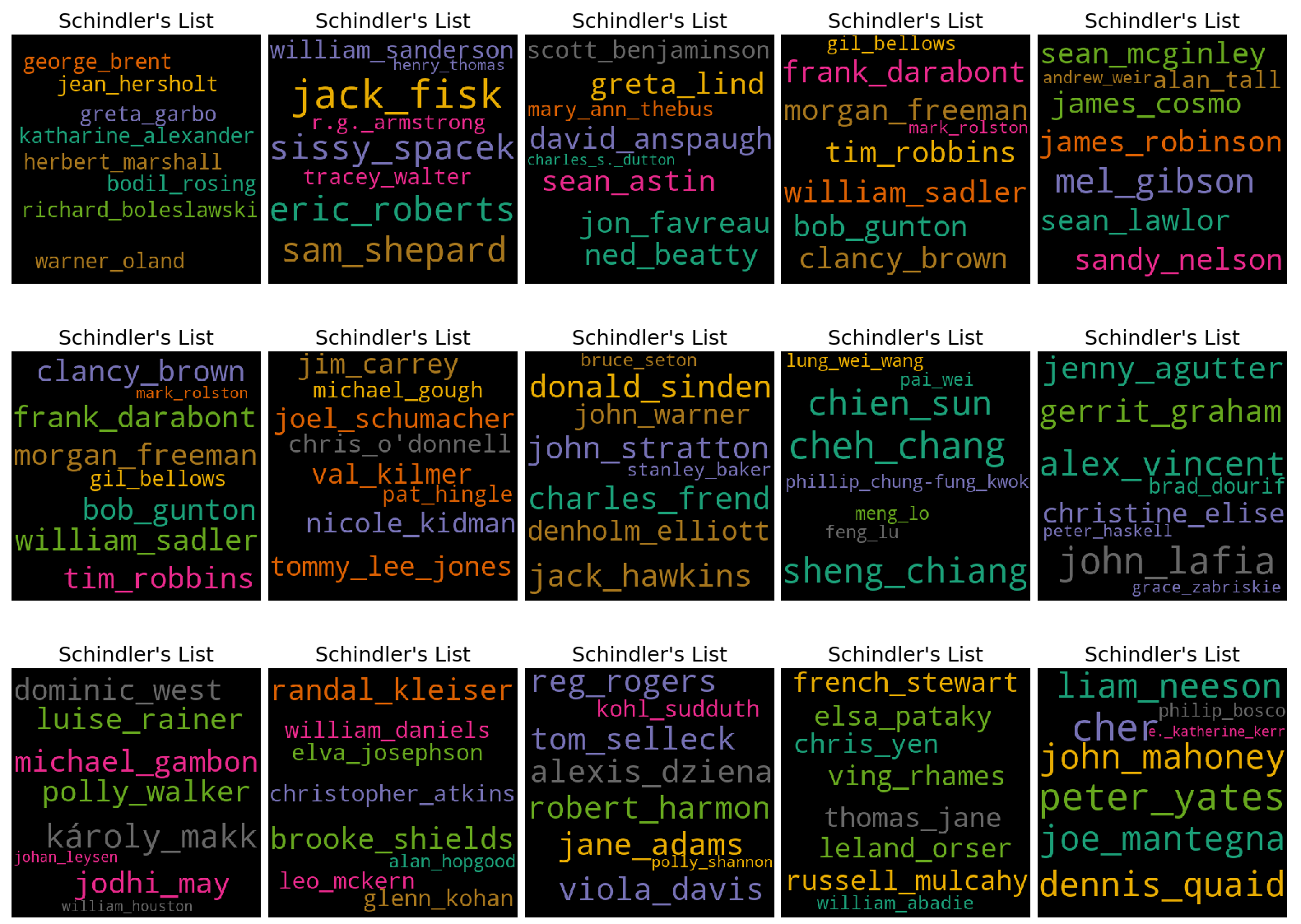}}
\hspace{-1\baselineskip}
\subfigure[]{\label{fig:a}\includegraphics[scale=0.25]{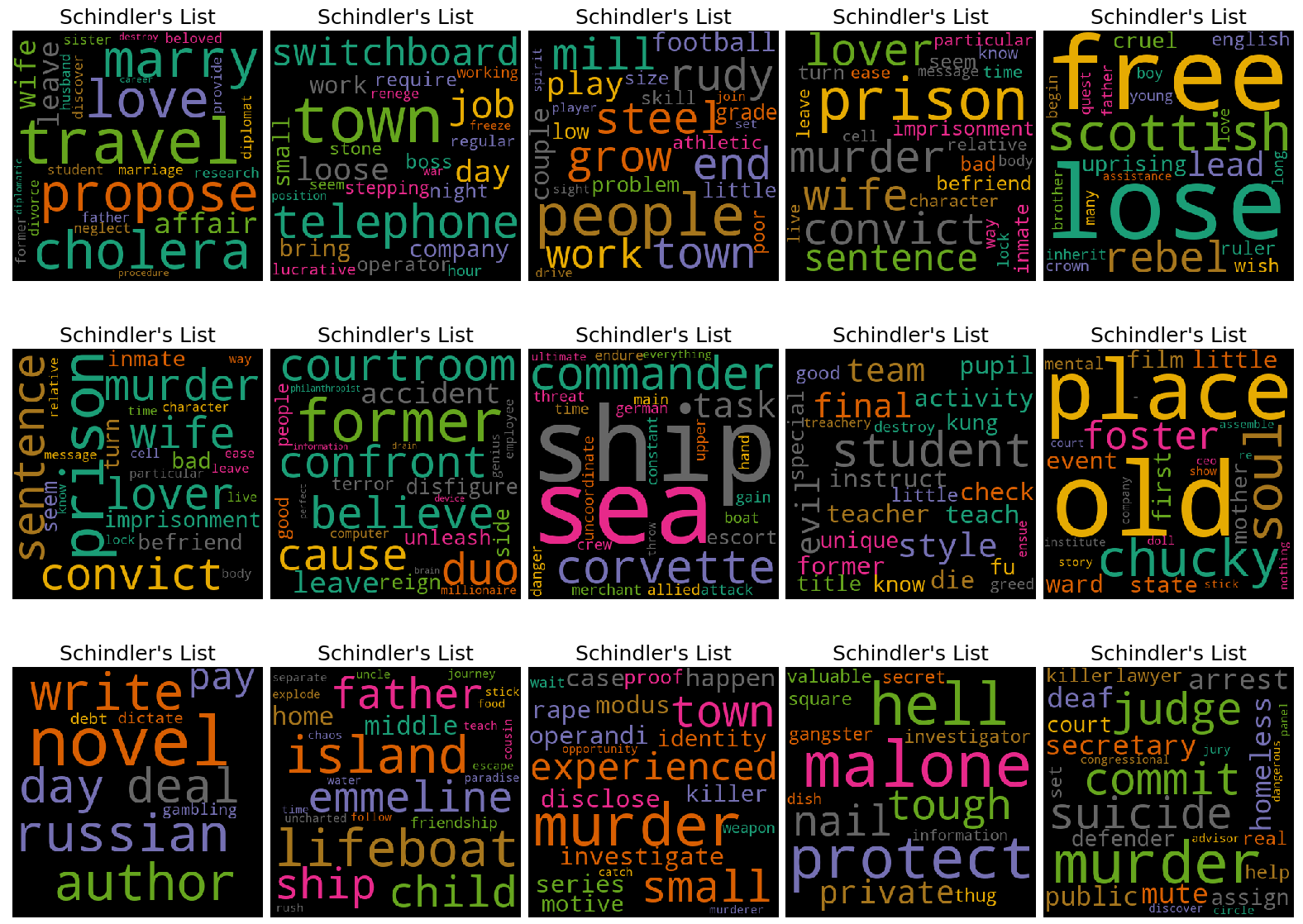}}
\hspace{-1\baselineskip}
\caption{Intepretability of cast and descriptions for individual users. Image a) shows tokens from move descriptions, and b) shows cast with the strongest positive attribution to output movies for 15 unique users. Token size reflects the attribution score. The recommended movie in all cases was "Schindler's List". The user ordering is the same as in Figure \ref{posters-individual}.}
\label{cast-desc-individual}
\end{figure*}

\begin{figure*}%
\centering
\includegraphics[scale=0.33]{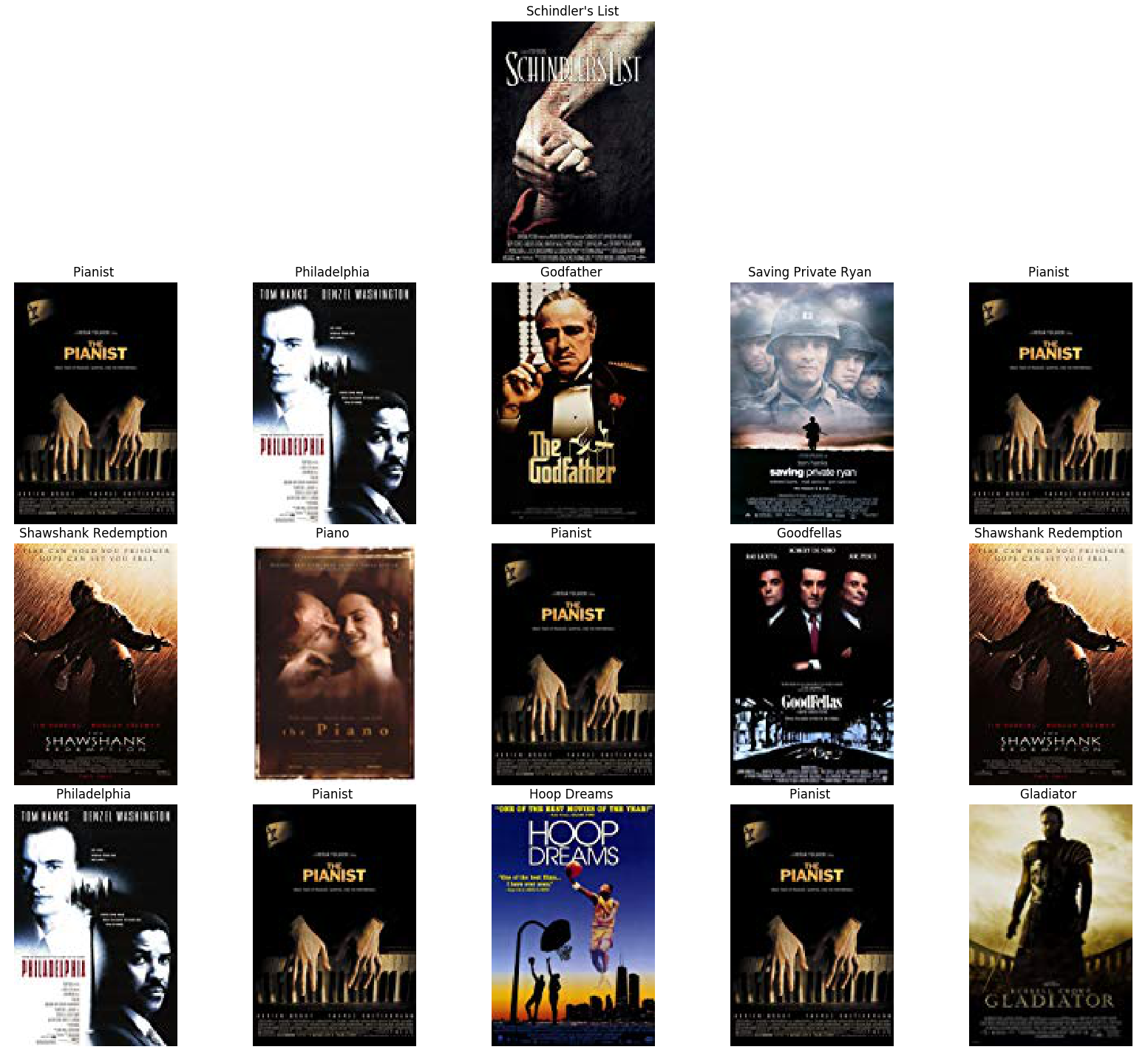}
\caption{Interpretability of posters for individual users. The top poster is the recommended movie poster ("Schindler's List), while the bottom posters are the most influential posters for this recommendation, for 15 unique users. The user ordering is the same as in Figure \ref{cast-desc-individual}.}
\label{posters-individual}
\end{figure*}


\section{EMDE Interpretability Method}

EMDE works by dividing the item space into regions and assigning items to specific buckets based on similarity of their embedding vectors. An encoded representation of a particular item is thus comprised of specific 'buckets' this item fell into, which are then combined into a joint item representation for each user following Count-Min Sketch algorithm \cite{CMS}. Such structures are used at both input and output of a simple feed forward neural network. To obtain approximate attributions for each input item, we first select the output buckets belonging to the item with the highest recommendation score. Then, we calculate attributions with Integrated Gradients \cite{10.5555/3305890.3306024}  just for these outputs. We sum the attributions at input, obtaining a flat vector of length identical to that of the input sketch: $n\_modalities \times n\_sketches \times sketch\_dim$. Note that at this step we do not yet have the attributions for each input items, as they are squashed into the input sketch representation. To obtain each individual input's attribution, we run a decoding procedure on the input sketch, similar to what is done at the prediction stage of EMDE: 1) Each item falls into $n$ buckets, where $n$ is defined by $n\_sketches$. For each input item, we identify the buckets into which the item fell originally, while creating the input sketches. 2) We average attribution values calculated for the buckets of an item; the resulting number is the individual item approximate attribution score. 3) We repeat this procedure for each modality.

\section{Experiments}
We train EMDE on MovieLens20M, a large-scale dataset of movie recommendations. After filtering done identically as in \cite{emde}, we obtain 20,108 movies and 116,677/10,000/10,000 train/test/valid users.
We use 5 modalities: liked movies interactions (for movies which received score of 4 or higher in the dataset), disliked movies interactions (for movies which received score lower than 4), movie cast, movie plot descriptions, and movie posters. As in \cite{emde} we encode the 4 modalities (liked movies, disliked movies, cast, and plots) using the EMDE simple embedding scheme.
We embed movie posters with SimCLR \cite{chen2020simple}. We follow \cite{emde} in the set up of the experiment. The output sketch is composed of just basic interactions and the input sketch consists of all 5 modalities.




Our evaluation procedure lets us discover most influential entities (e.g. most important tokens from movie plots) and the most influential input movies in terms of each modality (by aggregating the attribution scores for e.g. the tokens of each input movie descriptions). We find that supporting modalities often select a different movie, as the most influential, than the main modality - agreements range from 0\% (posters) to 10\% (movie plots and cast). This implies that supporting modalities do not just mirror the decisions of the main modality but also introduce additional, unique knowledge, even though their influence on standard performance measures is weak in this dataset \cite{emde}. Average attribution values for top influential movie calculated from each modality are comparatively high (ranging from 0.023 for cast to 0.026 for interactions) meaning that all modalities have a considerable influence on the output prediction.

We can either observe factors which are influential for a given movie in general (by aggregating scores across users), or per specific user. It is interesting to see that although the same movie might be predicted for several users, for each individual user it is influenced by different factors. It enables a comprehensive profiling of users.

We present our results of cast, description and poster interpretability in Figure \ref{cast-desc} and Figure \ref{posters}. Here the attributions are aggregated over users, which means that the tendencies shown are general. The ordering of movies is consistent across images, so comparison of different modalities is straightforward. In Figures \ref{cast-desc-individual} and \ref{posters-individual} we plot corresponding results for individual users.

\subsection{Intepretability of descriptions}
We find that high attribution scores often agree with common sense. For example, tokens which directly correspond to the plot of the recommended movie highly contribute to the prediction e.g. 'gang' and 'kill' are predictive of \textit{Fight Club} (Figure \ref{cast-desc}). However, often the most important tokens do not describe the plot but rather summarize the areas of interest for the movie's viewers. For example, tokens "know", "help", "see", "world" have high attribution scores for \textit{Gladiator}. These tokens do not seem to summarize the movie plot, which revolves around violence and fighting. Pure violence however seems ignored by this movie's viewers, who may pick this movie primarily because of its psychological drama features. In contrast, violence can be an advantage for viewers of \textit{Rear Window}, \textit{Fight Club}, and (surprisingly) \textit{Apollo 13}.

\subsection{Interpretability of cast}
Actors and directors often are strong influencers on other movies they have worked on. However, the contributions of particular cast members are not equal and we observe that some are way stronger than others. Sometimes the relationship proves more nuanced, which is exemplified in Figure \ref{cast-desc}. An actor who does not star in a movie but belongs to the same movie epoch can be a strong attribution - e. g. Joseph Cotten strongly contributes to the prediction of \textit{Casablanca}. Similarly, lead actors (even very famous ones) may not have a strong influence on prediction of their own movies. E.g. Russel Crowe is not a strong influencer for \textit{Gladiator} - instead, Will Smith is the decisive actor for this prediction, although he does not even star in this movie. If we rule out data and model bias, each such case might point to an important underlying phenomenon in the real world. For example, Russel Crowe may not have a large group of fans so dedicated as to go to his movies just because of his participation. Instead, his movies may be picked based on other factors, possibly more related to movie quality. In contrast, Alfred Hitchcock (\textit{Rear Window}) and Edward Norton (\textit{Fight Club}) might have such dedicated groups of followers. Of course, such hypotheses would demand a comprehensive analysis for each individual actor/director to be confirmed. However, the insights that follow can be very useful for star managers and movie producers alike. 

\subsection{Intepretability of posters}
With regard to posters, attributions let us rediscover the toolkit of movie poster creators (Figure \ref{posters}). The elements of the poster which are shared by the top output movie and the most influential movies present the visual characteristics which can be regarded as symbolizing the plot best. These elements often include the choice of color, similar positioning and number of characters, and spatial arrangement of items on the poster.

\subsection{Interpretability for individual users}
In Figure \ref{cast-desc-individual} and Figure \ref{posters-individual} we display the strongest influencers for the movie "Schindler's List" for 15 individual users. Again, the image ordering is consistent across the figures which enables cross-modal comparisons. It is evident that there is more variablity in the selection of most influential cast and description tokens, as compared to posters, which are often repetitive. This is yet another proof of the differences in behavior of each modality. Similarly, although the general tendencies of movie selection exemplified by the aggregate scores often agree with common sense, the scores for individual users show very varied reasons for the selection of a movie.

\section{Summary}
Our work shows that 1) the state-of-the-art recommender EMDE can be interpreted, and 2) EMDE gives many meaningful insights about preferences of moviegoers. Understanding the decision process behind recommendation systems can be helpful for the movie industry. 

\bibliography{example_paper}
\bibliographystyle{icml2020}

\end{document}